# Efficient coupling of light to graphene plasmons by compressing surface polaritons with tapered bulk materials

A. Yu. Nikitin[1,2], P. Alonso-González[1], R. Hillenbrand[1,2*]

1. CIC nanoGUNE Consolider, 20018 Donostia-San Sebastian, Spain
2. IKERBASQUE Basque Foundation for Science, 48011 Bilbao, Spain.
*Email: r.hillenband@nanogune.eu

**Graphene plasmons promise exciting nanophotonic and optoelectronic applications. Owing to their extremely short wavelengths, however, the efficient coupling of photons to graphene plasmons - critical for the development of future devices - can be challenging. Here, we propose and numerically demonstrate coupling between infrared photons and graphene plasmons by the compression of surface polaritons on tapered bulk slabs of both polar and doped- semiconductor materials. Propagation of the surface phonon polaritons (in SiC) and surface plasmon polaritons (in n-GaAs) along the tapered slabs compresses the polariton wavelengths from several micrometers to around 200 nm, which perfectly matches the wavelengths of graphene plasmons. The proposed coupling device allows for a 25% conversion of the incident photon energy into graphene plasmons and, therefore, could become an efficient route towards graphene plasmon circuitry.**

Graphene plasmons (GPs) are electromagnetic waves propagating along a graphene layer[1-10]. Their electrostatic tunability and extremely short wavelength, $\lambda_p$, being much smaller than the corresponding photon wavelength in free space, $\lambda_0$, promise exciting opto-electronic applications at the nanoscale. At infrared frequencies, GP wavelengths as short as 200 nm have been already observed experimentally[8,9]. Due to their extremely high momentum, GPs strongly concentrate electromagnetic energy, promising novel nanoscale photonic applications such as ultracompact tunable plasmonic absorbers[11,12], sensors, waveguides or modulators[10,13]. However, the huge momentum mismatch between free space photons and GPs challenges the efficient coupling between them.

Recently, it has been reported that propagating GPs can be excited by the strongly concentrated fields at the apex of metallic near-field probes[8,9]. The efficiency of the excitation mechanism, however, is unknown and expected to be low. Further, the experimental configuration of near-field microscopy does not constitute a practical solution for the development of integrated GP devices. The design of efficient couplers for GPs still remains a challenging task.

Several other configurations to overcome the momentum mismatch between the incoming light and GPs have been studied theoretically. For example, a periodic grating (placed for instance on top of a dielectric waveguide) with sufficiently short period $\Lambda$ can compensate the mismatch due to Bragg scattering in different diffraction orders $n$ following the expression $k_p = k_0 + nG$, with $G = 2\pi/\Lambda$ [14,15]. For efficient excitation of GPs, however, a large coupling length (in the order of the wavelength of the incident light) is required, which in turn requires a large propagation length of the GPs and prevents compactness of the device.

Another classical geometry common for the excitation of surface plasmons is the prism coupling method[16]. Taking into account the large GP momenta at mid IR and THz frequencies, the refractive index of the prism would need to be very large (~10-100), and thus practically not available.

We could also think of the excitation of GPs by compression of the photon wavelength in a dielectric waveguide with gradually increasing refractive index $n = n(x)$ and subsequent waveguide coupling to GPs, analogous to coupling plasmons in metal films to dielectric waveguides[17, 18]. However, compression of the mode wavelength in dielectric waveguides to the scale of GPs would again require materials with refractive indices up to 10 to 100 (as in the case of a prism coupler), which do not exist in nature at optical and infrared frequencies. Metamaterials could be used instead, which have been shown to support optical modes with ultrahigh refractive indices[19]. In this case, however, one faces an extremely challenging and highly demanding fabrication process.

Here, we introduce a novel concept (illustrated in Fig. 1) for the efficient launching of graphene plasmons based on the compression of surface polaritons (SPs) on tapered slabs of a bulk (3D) material supporting either surface phonon polaritons (SPhPs, for example on SiC slabs, or surface plasmon polaritons (SPPs, for example on highly doped n-GaAs slabs. The coupling mechanism is as follows: first, an incident propagating light beam is coupled via total internal reflection to an SP on a relatively thick slab (Otto prism coupling[20, 21])[22]. The thickness of the slab is chosen such that the SP wavelength can be easily matched by the photon wavelength in the prism, enabling high coupling efficiencies[20, 21]. Subsequent propagation of the SP along a tapered slab yields a compression of its wavelength. The slab thickness is gradually reduced until the SP wavelength matches that of the GPs. Finally, by placing the graphene above the thinned slab, the field of the SP couples efficiently to the GP.

We note that SPP compression on metal tapers has been already demonstrated at visible frequencies[23-26]. However, at mid-IR and THz frequencies, well below the plasmon resonance frequency, the compression of electromagnetic energy does not provide the necessary wavelength reduction, even if the mode diameter is compressed to below 100 nm[27, 28]. In order to achieve wavelength compression, the dielectric permittivity of the taper material, $\varepsilon_m$, needs to satisfy $\mathrm{Re}(\varepsilon_m) < -1$ (to support surface waves) and $|\mathrm{Re}(\varepsilon_m)| \gg |\mathrm{Im}(\varepsilon_m)|$, with $|\mathrm{Re}(\varepsilon_m)|$ being small (to provide sufficiently high momenta and low losses). Within the wide range of exciting plasmonic materials[29], highly doped semiconductors such as n-GaAs[30] and polar crystals such as SiC[31-33] fulfill these requirements in the infrared and mid-infrared spectral range, where GPs currently attract much interest[10, 11].

We first consider the dispersion relation of SPs in thin slabs of constant thickness d. For a slab with dielectric permittivity $\varepsilon_m$ embedded between two dielectric half-spaces with permittivities $\varepsilon_1$ and $\varepsilon_2$, the dispersion relation $k = k(\omega)$ follows from the solution of[21]

$$i \tan(q_{zm} k_0 d) = \frac{q_{zm}}{\varepsilon_m} \cdot \frac{q_{z1}\varepsilon_2 + q_{z2}\varepsilon_1}{q_{z1}q_{z2} + q_{zm}^2 \left(\varepsilon_1\varepsilon_2 / \varepsilon_m^2\right)}, \qquad (1)$$

where $k_0 = \omega/c$ is the momentum in free space, $q_{z,1,2,m} = \sqrt{\varepsilon_{1,2,m} - q^2}$, $q = k/k_0$ and $\text{Im}(q_{z,1,2,m}) \geq 0$. Eq. 1 can be derived from the Helmholtz equation, taking into account the boundary conditions for the fields at the interface between the slab and the dielectric half spaces. The solution of Eq. (1) yields symmetric and antisymmetric modes with respect to z-component of the electric field $E_z$[21]. In the following, we will consider only the antisymmetric solutions, as they exhibit higher momenta and are more sensitive to the slab thickness (compared to the symmetric solutions). For the substrate we choose $\varepsilon_2 = 2$ (i.e. a low refractive index material) and for the upper half space $\varepsilon_1 = 1$ (air).

In Fig 2a we show the dispersion for SiC slabs of different thickness $d$. It was obtained by solving numerically Eq. 1. The dielectric function of SiC was taken from Palik[34], and is given by

$$\varepsilon_m = \varepsilon_\infty \left( 1 + \frac{\omega_{LO}^2 - \omega_{TO}^2}{\omega_{TO}^2 - \omega^2 - i\omega\gamma} \right) \qquad (2)$$

with $\varepsilon_\infty = 6.56$ the background permittivity, and $\omega_{LO}$, $\omega_{TO}$ and $\gamma = 5.9$ cm$^{-1}$ the transversal and longitudinal phonon frequencies and relaxation rate, respectively. Between the transversal and longitudinal phonon frequencies, $\omega_{TO} = 800$ and $\omega_{LO} = 970$ cm$^{-1}$, the dielectric function of SiC is negative, which indicates that SiC slabs support SPhP in this spectral range. For all thicknesses d, we see the typical dispersion for SPhPs. The wavevector $k$ increases with frequency $\omega$, reaching a maximum at approximately the surface phonon resonance at $\text{Re}(\varepsilon_m) = -1$ ($\omega = 950$ cm$^{-1}$). Notice that a double-humped behavior near phonon resonance is due to nonsymmetric dielectric surrounding of the film. At each hump the mode is localized at one of the film faces (compare the positions of the humps with the shown in Fig 2 (a) dispersion curves for the semi-infinite SiC bounding either to air or to the substrate). For slabs thicker than 300 nm, the dispersion is similar to that of SPhPs on bulk SiC surfaces, with wavevectors $k$ being not more than a factor of 2 to 5 larger than the corresponding photon wavevector $k_0$. With decreasing $d$, we find a dramatic increase of the SPhP wavevector, which can be 10 to 50 times larger than $k_0$. Fig. 2b shows the mode profiles (real part of the electric field) for $\omega = 890 \,\text{cm}^{-1}$ (indicated by a dashed light gray curve in Fig. 2a), illustrating the decreasing SPhP wavelength with decreasing thickness $d$. For slabs as thick as 50 nm, the SPhP wavelength is already 20 times smaller than the photon wavelength. Analogous to SPPs on metals, the SPhP wavelength reduction is accompanied by a huge field confinement as clearly observed in Fig. 2b., but also by strong damping. The propagation length, however, is still in the order of several SPhP wavelengths . Most intriguing, the SPhPs wavevectors on 50 nm thick SiC slabs can be as large as the GPs wavevectors, and between 890 and 920 cm$^{-1}$ can even exceed them. We show this finding by plotting in Fig. 2a the dispersion of GPs (dashed line, see details in the next paragraph) in a free-standing doped graphene layer. As an example we have chosen a Fermi energy $|E_F| = 0.44 \,\text{eV}$, for which mid-infrared graphene plasmons have been already observed experimentally[8].

The GP dispersion for free-standing graphene in the IR region, where the GP momentum is large, reads[1]

$$q_{GP} \simeq i/\alpha \quad \text{with} \quad \alpha = 2\pi\sigma/c \quad (2)$$

where $\sigma$ is the optical conductivity calculated within the local random phase approximation[3, 35, 36]. Interestingly, for large SPhP momenta and vanishing slab thickness ($qdk_0 \ll 1$), the dispersion relation (1) greatly simplifies to

$$q \approx i/\alpha_{eff} \quad \text{with} \quad \alpha_{eff} = ik_0 d \frac{\varepsilon_m + \varepsilon_1\varepsilon_2/\varepsilon_m}{\varepsilon_1 + \varepsilon_2} \quad (3)$$

This shows that SPhP in a very thin 3D SiC slab have the same dispersion as GPs (2), with the effective normalized conductivity $\alpha_{eff}$ proportional to the thickness of the layer $d$. We would also like to point out the recent experimental observation of SPs in thin slabs of hBN[37].

SPs with extraordinarily high wavevectors - similar to GPs - can be also supported by thin slabs of a doped semiconductor material, as we demonstrate in Figs. 2c,d for the case of n-doped GaAs with a carrier density of n=$10^{20}$ cm$^{-3}$ and mobility of n=$10^3$ cm$^2$V$^{-1}$s$^{-1}$ [29]. The SPPs dispersion in doped semiconductors can be calculated following the eq. 1 with the dielectric function $\varepsilon_m$ described by the Drude-Lorentz model,

$$\varepsilon_m = \varepsilon_b - \frac{\omega_p^2}{\omega(\omega + i\gamma)} \quad (4)$$

with $\varepsilon_b = 10.91$ the background permittivity, and $\omega_p = 11485$ and $\gamma = 137$ cm$^{-1}$ the plasma frequency and relaxation rate, respectively. The plasma frequency $\omega_p \sim \sqrt{n}$ is determined by the mobile carrier concentration in the semiconductor $n$, which can be adjusted by the degree of doping, thus providing a tuning capability for the SPP dispersion. Analogously to SPhPs on thin SiC slabs, we observe in Figs. 2c,d SPPs on thin n-doped GaAs with large momentum, reaching values well above the momentum of GPs. This can be appreciated by comparing the SPP dispersion with the GP dispersion on graphene with a Fermi energy of 0.8 eV (dashed line in Fig. 2c).

Altogether, Fig. 2 clearly shows that thin slabs of a bulk material can support SPs with wavelengths comparable to that of GPs in the infrared and mid-infrared spectral regions. While GPs may offer better tunability and simpler fabrication procedures, thin slabs of a bulk material offer the possibility to adjust the SP wavelength and mode confinement at a given frequency over a much larger range, from nearly the wavelength of incident photons to the much smaller wavelength of GPs and beyond. Such wavelength adjustment can be accomplished by simply reducing the slab thickness from a few hundreds of nm to a few tens of nm. In particular, the wavelength reduction offers the possibility to couple photons to GPs by propagating SPs along tapered slabs. In Fig. 3 we demonstrate the compression of SPhPs and SPPs on tapered SiC (Fig. 3a,b) and n-GaAs (Fig. 3c,d) slabs, respectively. The frequency for the case of SiC is 889 cm$^{-1}$ while for n-GaAs it is 2016 cm$^{-1}$. We show the real part of the vertical electric field along the tapered slab (thickness reduces from left to right) in Figs. 3a,c and the corresponding absolute value of the field in Fig. 3b,d. We clearly see that both the mode wavelength and mode volume decrease in the propagation direction (from left to right), which in turn induces a pronounced intensity enhancement at the end of the taper. For the SPhPs on the SiC taper, the mode index increases from $q_{in} = 2.62 + 0.14i$ (400 nm slab thickness) to $q_{out} = 17.4 + 1.25i$ (50 nm slab thickness), while for n-GaAs taper we find $q_{in} = 1.76 + 0.069i$ (100 nm slab thickness) and $q_{out} = 21.2 + 2.26i$ (5 nm slab thickness). At the end of the taper, the SPs continue their propagation over several wavelengths along a 50 nm thick SiC slab and a 5 nm thick GaAs slab, respectively. However, the intensity decays significantly within a few mode wavelengths. Due to the subwavelength-scale field confinement, a large part of the electromagnetic energy propagates inside the slab, thus damping the mode. The mode propagation length in units of the SP wavelength (figure of merit) is given by $L = \text{Re}(k)/[2\pi \text{Im}(k)]$, yielding $L_{in} = 2.98$ at the beginning of the SiC taper (400 nm slab thickness) and $L_{out} = 2.21$ at its end (50 nm slab thickness). Remarkably, the strong mode compression along the taper does not significantly reduce L, highlighting that SiC is an interesting low-loss polaritonic material. Note that the strong reduction of the absolute propagation length is mainly due to the dramatic reduction of the mode wavelength. For the GaAs taper we find $L_{in} = 4.06$ (100 nm slab thickness) and $L_{out} = 1.49$ (5 nm slab thickness). In contrast to SiC, the propagation length relative to the mode wavelength diminishes clearly.

We finally demonstrate in Fig. 4 how the compression of SPs on tapered slabs can be employed for the efficient launching of GPs. We consider in the following a tapered SiC slab, because of the relatively weak losses during the mode compression. In order to achieve effective coupling to GPs, the graphene sheet is placed at a distance of 150 nm above the thin SiC slab extending from the taper. From the point of view of fabrication, this could be accomplished by depositing a thin dielectric layer onto the SiC slab (for instance $CaF_2$ or $SiO_2$) and then the graphene on top of this layer (see Fig. 1). As the optical properties of the dielectric layer will not strongly affect the coupling (provided that the dielectric permittivity of the layer $\varepsilon_d$ is small), we consider in the following $\varepsilon_d = 1$, corresponding to free-standing graphene.

In Figs. 4a and b we show colorplots for both the real part of the vertical field and the absolute value of the total field along the coupling device, $\mathrm{Re}(E_z)$ and $|\mathbf{E}|$. The tapered SiC slab and parameters are the same as in Figs. 3a,b. For the graphene layer we assume a Fermi energy of $|E_F| = 0.44$ eV, which is close to typical values observed in CVD grown graphene[8]. For the mobility we assume $11.36 \cdot 10^3$ cm$^2$/(V•s) (corresponding to 0.5 ps relaxation time of the charge carriers). This value is elevated compared to graphene samples of recent plasmonic studies[8,9], but it allows for better illustration of the coupling between the slab SPs and GPs. On the other hand, such mobilities seem to be available in the future with graphene samples of improved quality[38,39].

Figs. 4a and b demonstrate the coupling between compressed surface phonon polaritons on a thin SiC slab and a graphene layer, both being infinitely long (right side of the figures). We observe that the field intensity (Fig. 4b) is periodically transferred from the SiC slab to the graphene and vice versa. Such behavior is well known from waveguide couplers[18,40-42] and originates from the beating between the two modes in closely spaced waveguides. In our case, the beating modes originate from the coupling of plasmons in the graphene sheet and the SPhPs in the SiC slab.

According to Fig 4c, the beating period is $L_b = 1.4$ μm. This can be compared with the results of coupling mode theory[40], according to which $L_b = 2\pi/|\text{Re}(k_1 - k_2)|$ where $k_1$ and $k_2$ are the wavevectors of the modes of the waveguide coupler. From the mode analysis (not shown here) of the coupled infinite graphene sheet and 50 nm thick SiC slab we find $k_1/k_0 = 19.22 + 0.39i$ and $k_2/k_0 = 11.27 + 0.76i$, so that $L_b = 1.43$ μm, which is in an excellent agreement with the value found from the first principle calculations (Fig 4b). Notice that the length at which the maximum energy transfer occurs is $L_b/2$.

By terminating the SiC slab at the position of the maximum field in graphene (marked by arrows in Figs. 4a and b), the transferred SPhPs energy continues its propagation as plasmon in the graphene sheet (Figs. 4c and d). In the presented example, the thin SiC slab is terminated at a distance $x_c$ ; $L_b/2 = 700$ nm from the taper end (see also schematics in Fig. 5). We note that the coupling mechanism (after the mode compression) is similar to that of dielectric photonic modes and surface plasmons in metal slabs[17, 18, 41].

Figs. 4c and d indicate already an efficient excitation of propagating GPs within a coupling length $L_c = 600$ nm, which is well below the photon wavelength ($\lambda_0 = 11.4$ μm). This represent a significant advantage compared, for instance, to grating couplers on dielectric waveguides, where the coupling length needs to be larger than one photon wavelength[15]. For that reason, even GPs with short propagation lengths can be efficiently excited.

Due to the beating, the energy from the SPhP can be almost totally transferred to GPs (see below and Supporting Information). However, the total efficiency for converting photons into GPs is smaller, owing to losses (absorption, out-of-plane scattering and back-reflection) during the compression of the SPhPs along the taper. Further reduction of the coupling efficiency might be caused during the coupling of photons to the initial SPhP mode on the thick slab. By proper design of a prism or grating coupler[21], photons might be converted with high efficiency into the initial SPhP mode on the 400 nm thick SiC slab, owing to its rather large wavelength of 5.25 μm ( $= 0.46 \lambda_0$ ). In the following, therefore, for a numerical study of the coupling efficiency we only consider the energy transfer from the initial SPhP mode to the GP mode in the configuration shown in Figs. 4c and d. To that end, we introduce the efficiency $\eta$ as the ratio of the energy absorbed in the graphene $A$ (which is essentially dominated by GPs) to the Pointing vector flux of the incident polaritonic mode at the entrance $P_0$, i.e. $\eta = A/P_0$. To be free of any particular realization of the device we take the entrance of the initial SPhP mode at the beginning of the taper.

In Fig 5(a) we show the efficiency $\eta$ as a function of the tapering angle $\theta$ for two different frequencies and different final thicknesses of the SiC slab: $\omega = 877$ cm$^{-1}$ and $d_2 = 50$ nm (red curve), and $\omega = 855$ cm$^{-1}$ and $d_2 = 25$ nm (blue curve). The initial thickness of the slab is $d_1 = 400$ in both cases. The length of the final slab, $x_c = 700$ nm (which is approximately a half of a beating period $x_c$ ; $L_b/2$), has been chosen such that the termination of the slab coincides with the position of maximum energy transfer into the graphene layer (see Fig. 4(b)). We find the maximum efficiency for taper angles $\theta = $ 15 and 11.5 degrees, respectively, which we explain by the compromise between field absorption in the taper (increasing with increasing taper length, i.e. smaller taper angle) and back reflections at and inside the taper (increasing with increasing taper angle, owing to an increasing impedance mismatch).

In Fig 5b we study how the efficiencies depend on the frequency. Each spectrum shows a resonance peak, which appears at the frequency where the SPhP and GP dispersion curves cross, i.e. where the momentum of the SPhP best matches the GP momentum (see Fig. 2a). Note that the curves in Fig. 5a were calculated at the frequencies where the efficiency reaches its maximum (877 cm$^{-1}$ for the red curve and $\omega = 855$ cm$^{-1}$ for blue curve, respectively).

According to the results presented in Fig. 5, the maximum efficiency of the coupler is higher than 25% at a frequency $\omega = 877$ cm$^{-1}$. This is a rather large value, especially when considering that the coupling length is comparable to the GP wavelength and that the SP mode is progressively absorbed during its propagation along the taper. To better understand the coupling losses, we evaluated the coupling efficiency independent of the taper (see Supporting Information). To that end, we calculated the coupling of SPhPs in an untapered 50 nm-thick SiC slab (corresponding to the thinnest part of the taper) to GPs. We found a coupling efficiency of about 70% at $\omega = 877$ cm$^{-1}$. This means that the main amount of losses are due the SPhP compression in the taper, and not due losses during the coupling process.

In conclusion, we have demonstrated that by propagating surface phonon or plasmon polaritons along tapered slabs of a bulk (3D) polaritonic material, it is possible to efficiently launch graphene plasmons with a predicted efficiency of more than 25 %.

We further stress that thin slabs of bulk (3D) polaritonic materials – by themselves - could become an interesting platform for the development of short-wavelength surface polariton photonics for sensing and communication applications, particularly in the infrared spectral range where phonons and plasmons in materials such as SiC or n-GaAs exhibit low losses compared to plasmonic materials typically used at visible and near-IR frequencies (e.g. gold).

**Acknowledgements**

This work was financially supported by the ERC Starting Grant No. 258461 (TERATOMO) and EC under Graphene Flagship (contract no. CNECT-ICT-604391). AYN acknowledges the Spanish Ministry of Science and Innovation grant MAT2011-28581-C02.

**Methods**

The calculations have been performed with the help of the finite elements methods using Comsol software. In order to simulate the propagation of the polaritonic mode, the profile of the mode in the analytical form has been set in the boundary conditions for the left boundary of the rectangular domain. The graphene layer has been modeled as a surface current in the boundary conditions. The domain has been discretized by using an inhomogeneous free-triangulat mesh with the maximal element size being less than 5% of the local spacial oscillation period of the field. The convergence has been assured both for electromagnetic fields and power fluxes (see Supporting Information).

**Supporting Information**

Description of additional details of the calculations. This material is available free of charge via the Internet at http://pubs.acs.org.

**Figures**

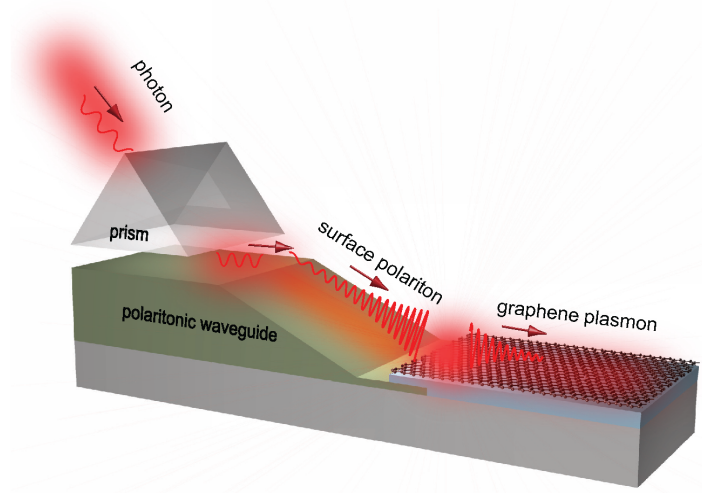

**Figure 1:** Schematic for the full realization of the proposed device.

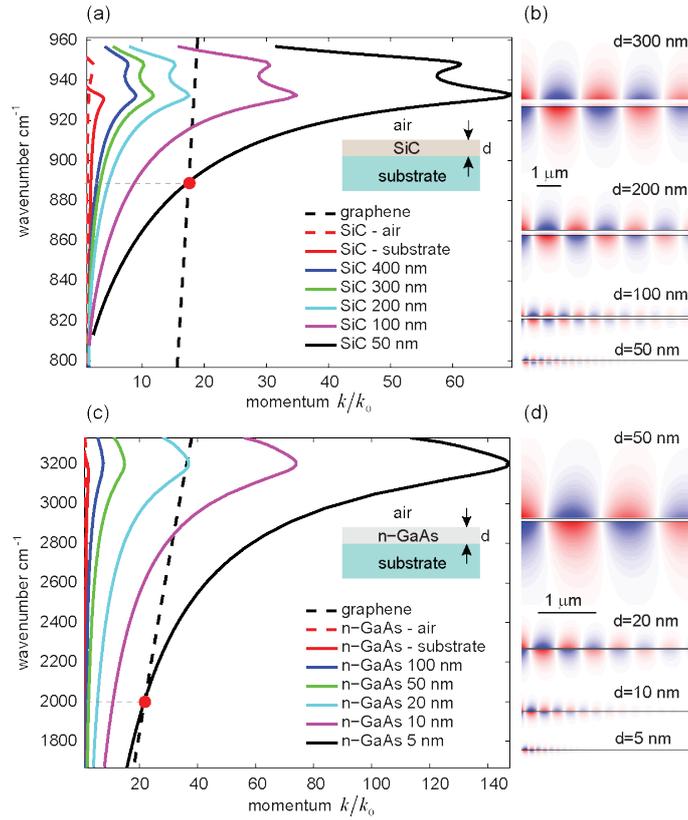

**Figure 2:** Dispersion of the surface polaritons in slabs placed on a substrate and the dispersion of the graphene plasmons. a) Slabs of SiC. The Fermi level of graphene is | $|E_F| = 0.44\,\text{eV}$. c) Slabs of n-GaAs. The carrier concentration of the semiconductor is $10^{20}$ cm$^{-3}$, the rest of the parameters are taken from[29]. The Fermi level of graphene is $|E_F| = 0.8\,\text{eV}$. b,d) Snapshots of the vertical electric field $\text{Re}(E_z)$ for different thicknesses of the slabs $d$. For comparison, the dispersion curves for the semi-infinite polaritonic media bounding either to air or to the substrate are shown in both panels (continuous and discontinuous red curves). The refractive index of the substrate is $\varepsilon_2 = 2$ in all cases.

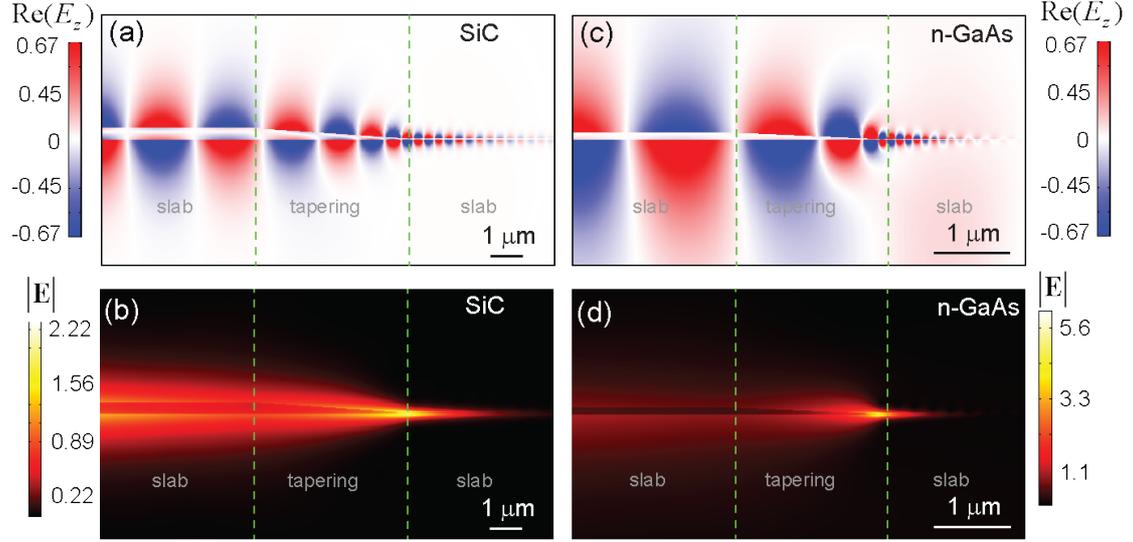

**Figure 3:** Snapshots of the real part of the vertical component of the electric field $\text{Re}(E_z)$ and the spatial distribution of the absolute value of the electric field $|\mathbf{E}|$ for the modes propagating inside the tapered waveguides. a,b) SiC waveguide with an initial and a final thickness of 400 and 50 nm, respectively; the tapering angle is 4 deg. The frequency is 889 cm$^{-1}$. c,d) n-GaAs waveguide with an initial and a final thickness of 100 and 5 nm, respectively; the tapering angle is 2.7 deg. The frequency is 2016 cm$^{-1}$. The fields are normalized to the maximum value of $|\mathbf{E}|$ of the incoming mode. The refractive index of the substrate is $\varepsilon_2 = 2$, as in Fig. 2.

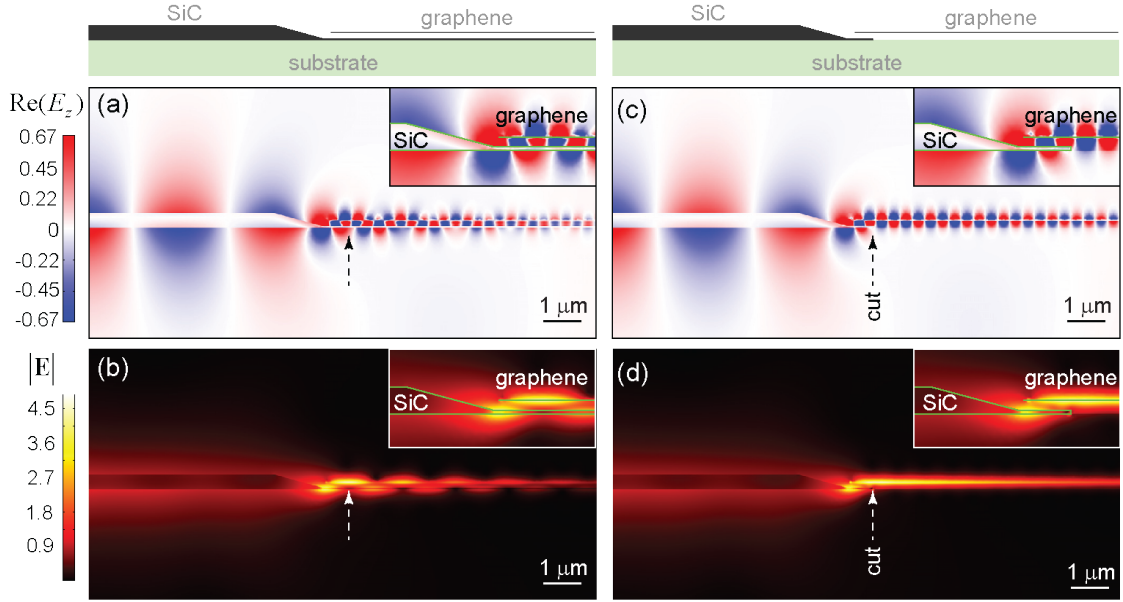

**Figure 4:** Snapshots of $\mathrm{Re}(E_z)$ and the spatial distribution of $|\mathbf{E}|$ for the tapered SiC waveguide attached to the graphene sheet. The tapering angle is 15 deg, its length 1.3 µm. The initial and final thicknesses of SiC slab are 400 and 50 nm, respectively. The graphene-SiC vertical separation is 150 nm. a,b) Infinite waveguide. c,d) Cut waveguide. The length of the thinnest part is 700 nm. The SiC-graphene overlap along the x-axis is 600 nm. The frequency is 877 cm$^{-1}$. The substrate and the parameters of graphene are the same as in Fig. 2. The fields are normalized to the maximum value of $|\mathbf{E}|$ of the incoming mode. The upper schematics show the geometry of the structures. The insets present the zoomed-in regions of the taper.

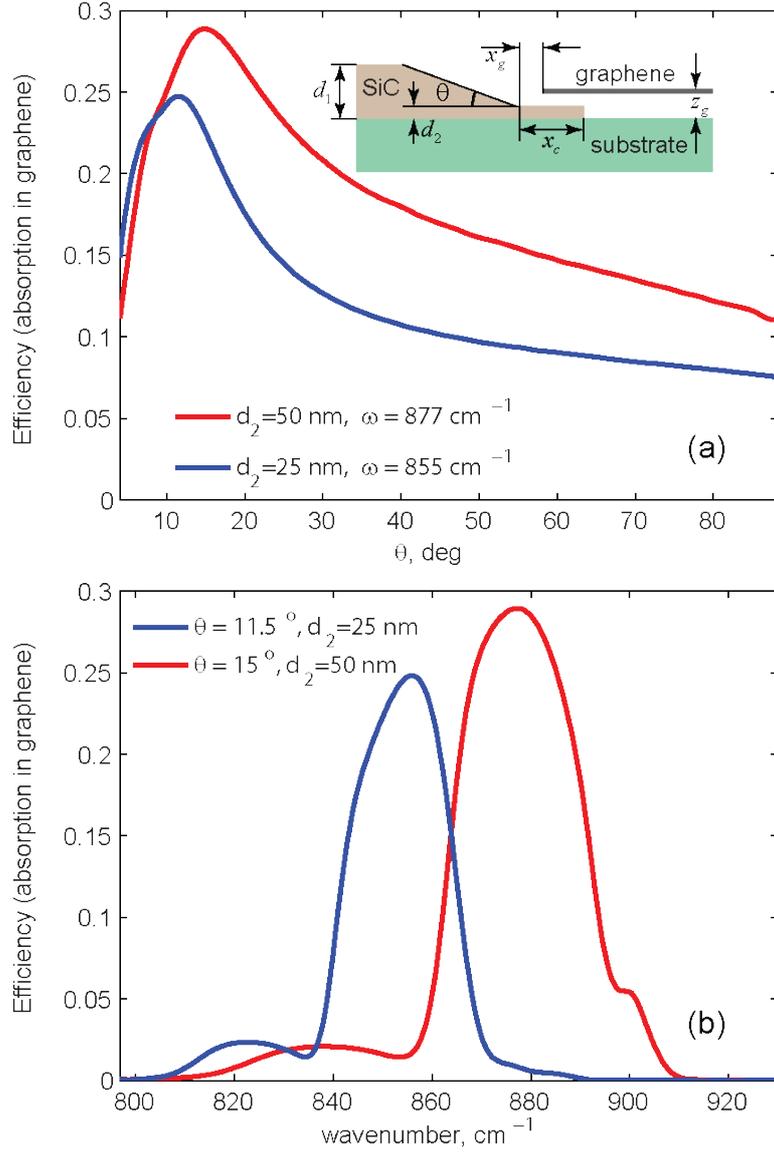

**Figure 5:** Coupling efficiency of the cut tapered SiC waveguide. The initial thickness of the waveguide is $d_1 = 400$ nm. The length of the thickest slab is 5 μm. The graphene vertical and horizontal separations are $z_g = 150$ nm and $x_g = 100$ nm respectively. a) The efficiency as a function of the tapering angle $\theta$ for two different final slab thicknesses $d_2$ and frequencies. The red curve is for $\omega = 877$ cm$^{-1}$ and $d_2 = 50$ nm, while the blue curve is for $\omega = 855$ cm$^{-1}$ and $d_2 = 25$ nm. b) The efficiency as a function of frequency for two different values of $d_2$ and angles $\theta$. The length of the thinnest part is $x_c = 700$ nm in both configurations.

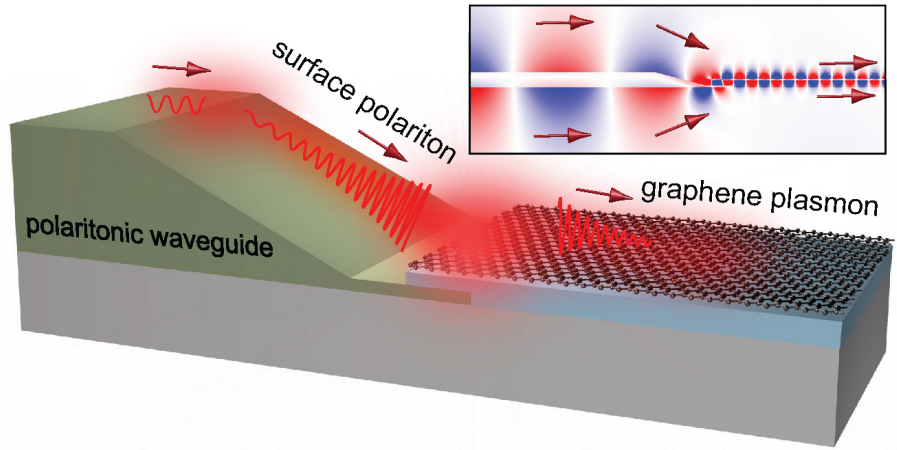

**Figure TOC**